# Multiferroicity and hydrogen-bond ordering in $(C_2H_5NH_3)_2CuCl_4$ featuring dominant ferromagnetic interactions

B. Kundys,[1,*] A. Lappas,[2] M. Viret,[1] V. Kapustianyk,[3,4,5] V. Rudyk,[3,4] S. Semak,[4,5] Ch. Simon,[6] and I. Bakaimi[2,7]

[1]*Service de Physique de l'Etat Condensé, DSM/IRAMIS/SPEC, CEA Saclay URA CNRS 2464, 91191 Gif-Sur-Yvette Cedex, France*
[2]*Institute of Electronic Structure and Laser, Foundation for Research and Technology, P.O. Box 1385, Vassilika Vouton, 711 10 Heraklion, Crete, Greece*
[3]*Scientific-Technical and Educational Center of Low-Temperature Studies, Ivan Franko National University of Lviv, Dragomanova str. 50, UA-79005 Lviv, Ukraine*
[4]*Scientific and Educational Center "Fractal," Ivan Franko National University of Lviv, Dragomanova str. 50, UA-79005 Lviv, Ukraine*
[5]*Physical Department, Ivan Franko National University of Lviv, Dragomanova str. 50, UA-79005 Lviv, Ukraine*
[6]*Laboratoire CRISMAT, CNRS UMR 6508, ENSICAEN, 6 Boulevard du Maréchal Juin, 14050 Caen Cedex, France*
[7]*Department of Physics, University of Crete, P.O. Box 2208, 71003 Heraklion, Greece*

We demonstrate that ethylammonium copper chloride, $(C_2H_5NH_3)_2CuCl_4$, a member of the hybrid perovskite family is an electrically polar and magnetic compound with dielectric anomaly around the Curie point (247 K). We have found large spontaneous electric polarization below this point accompanied with a color change in the sample. The system is also ferroelectric, with large remnant polarization (37 $\mu C/cm^2$) that is comparable to classical ferroelectric compounds. The results are ascribed to hydrogen-bond ordering of the organic chains. The coexistence of ferroelectricity and dominant ferromagnetic interactions allows to relate the sample to a rare group of magnetic multiferroic compounds. In such hybrid perovskites the underlying hydrogen bonding of easily tunable organic building blocks in combination with the 3$d$ transition-metal layers offers an emerging pathway to engineer multifuctional multiferroics.

Vigorous research efforts on magnetoelectric effects in the recent years[1–3] has stimulated the exploratory design and synthesis of magnetoelectric materials[4] and reinvestigation of known compounds, where $BiFeO_3$ stands out as a prominent example.[5] Single phase materials where magnetism and ferroelectricity (FE) coexist are rare. This is basically due to the mutual exclusive requirement for closed-shell cations for the former and unpaired spins on the same lattice site for the latter mechanism to be operative. Broadly speaking there are two families[6] of multiferroics (i) materials where electric polarization and magnetism have different origins and consequently different ordering temperatures due to the different functional subsystems and (ii) materials where the polar state takes place only when the specific (antiferro)magnetically ordered state is established. In the latter case, the spin driven microscopic mechanisms[7,8] (including spontaneous magnetostriction[9,10]) can produce the loss of inversion symmetry at the spin reorientation transition and give rise to an electrically polar state.

Although the strongest magnetoelectric interactions are expected between ferromagnetic (FM) and ferroelectric subsystems the majority of the investigated magnetoelectric materials today are antiferromagnetic (AFM). In that respect identifying a ferroelectric ferromagnet or magnetic material, with dominant ferromagnetic interactions, as reported here, is indeed a demanding task for the modern solid-state materials physics. The early discovery of the weakly ferromagnetic ($T_c$=60 K) but also electrically polarizable nickel-boracite $Ni_3B_7O_{13}I$ system,[11] where the three-dimensionally linked metal-halogen-metal chains can support not only collective magnetic phenomena but also their interaction with electric polarization, has not been followed yet by other similar successes. The aforementioned compound belongs to a not so well exploited class of materials involving functional building groups, with strong dielectric response, embedded though in a sublattice of magnetic cations; this conceptually results into two subsystems within a single phase chemical compound. In view of this idea Cui *et al.*[12] have recently reported that the porous molecular crystal $[Mn_3(HCOO)_6](C_2H_5OH)$, where ethanol are guest molecules, can display FE at 163 K and a ferrimagnetic transition at lower temperature (8.5 K). Their results suggest the significant role of the interaction between host lattice and the guest[13] as a means to the rearrangement of guest species and concomitantly to a transition in the dielectric response.

Host-guest interactions alone and/or mechanisms involving hydrogen-bond ordering schemes, responsible for generating electric polarization in nonmagnetic systems are known in the scientific community (cf. potassium dihydrogen phosphate goes from paraelectric to ferroelectric at 122 K).[14,15] It is postulated that order-disorder phase transitions, rather than displacive ones, intrinsic to the broad family of hybrid organic-inorganic compounds can act as a good alternative to engineer multiferroic behavior in such previously unexplored materials. A way forward was recently shown by the exciting opportunity opened by metal-organic frameworks (MOFs).[16] Beside their unique potential in hydrogen storage, catalysis, separation techniques, nonlinear optical properties, etc, these materials can also become multiferroic. One of such possibilities is shown by the $[(CH_3)_3NH_2]Mn(HCOO)_3$ MOF, with a perovskite-like topology. A change in the molecular structure at 185 K, from a disordered to a more ordered state due to the existence of hydrogen bonding between the MOFs components, causes a paraelectric to antiferroelectric phase

KUNDYS *et al.*

transition that coexists with weak ferromagnetism upon further cooling (8.5 K).[17] These early results demonstrate that it is worth searching for multiferroic behavior within materials where electrical order involves hydrogen bonding. This pathway can be especially rewarding if the chemistry of such compounds can be readily optimized to allow compositions supporting the above-mentioned cooperative phenomena close to room temperature.

In that respect revisiting previously known hybrid crystal structures comprising of metal ions in complex with organic molecules brings up an attractive opportunity to explore their magnetoelectric properties that are highly promising for information storage[18] or electric field controlled magnetic sensors. A good candidate to meet this challenge is the family of compounds with composition $(C_nH_{2n+1}NH_3)_2MCl_4$, where $M$ is a divalent metal ($M=Mn^{2+}$, $Cd^{2+}$, $Fe^{2+}$, and $Cu^{2+}$). This series crystallizes in the layered perovskite structure,[19–22] consisting of infinite, staggered layers of corner-sharing $MCl_6$ octahedra interleaved by alkylammonium cations $[C_nH_{2n+1}NH_3^+$; abbreviated as MA for methylammonium ($n=1$), EA for ethylammonium ($n=2$), etc.]. The cavities between octahedra are occupied by the ammonium heads of the organic cations, which importantly form strong N-H···Cl hydrogen bonds to any of the eight halides. In this structure, adjacent layers are stacked upon one another through van der Waals force between terminal alkyl groups. It is worth noting that the choice of the hydrogen-bonding scheme is important for determining the orientation and conformation of the organic molecules within the layered hybrid structure and in effect can influence the temperature evolution of the structural phase transitions in the perovskite structure.[23–25]

Here we report that multiferroic behavior is indeed possible in a representative member of this vast family of the organic-inorganic perovskites, highlighting the hydrogen-bond ordering scheme as an emerging route toward multifunctional multiferroics. The system of interest, the $(C_2H_5NH_3)_2CuCl_4$ (hereafter $EA_2CuCl_4$) compound, was grown in a single-crystal form at room temperature by slow evaporation of an aqueous solution of $CuCl_2$ and $C_2H_5NH_3Cl$ salts taken in the stoichiometric ratio. This system crystallizes in an orthorhombic cell (space group, $Pbca$; $a=21.18$ Å $b=7.47$ Å, and $c=7.35$ Å) (Ref. 19) at room temperature. It is comprised of Jahn-Teller (J-T) distorted $Cu^{2+}$ layers separated by two layers of ethylammonium groups (Fig. 1). Notably, this compound undergoes a series of structural phase transitions[26] as a consequence of modifications not only due to the J-T effect but importantly due to the arrangement (orientational order and conformation of the organic molecules) of the ethylammonium chains.[27,28] With temperature increase these entail: triclinic ($T_4=232$ K)$\rightarrow Pbca$ ($T_3=330$ K)$\rightarrow$rhombic ($T_2=356$ K) $\rightarrow P2_1/c$ ($T_1=364$ K)$\rightarrow Bbcm$.

Magnetization was measured in a superconducting quantum interference device magnetometer with the magnetic field ($H$) applied parallel to the metal-halogen layers. The measurements were undertaken while cooling the sample down in a field of 10 Oe (i.e., a field-cooled protocol). Dielectric measurements were performed in the cryostat of a PPMS by using an Agilent 4248A RLC bridge (at 100 mV oscillation voltage amplitude) while polarization was mea-sured with a Keithley 6517A electrometer. In the latter the electric field ($E$) was applied perpendicular to the ionic layers of $CuCl_4^{2-}$. The ferroelectric loop was measured by the modified dc ferroelectric test method,[29] employed before to identify ferroelectricity in magnetic multiferroics.[30–32] For both polarization and dielectric measurements fine copper electrodes were attached on the parallel sides of the platelike (typically $2\times3\times0.4$ mm$^3$) $EA_2CuCl_4$ single crystals.

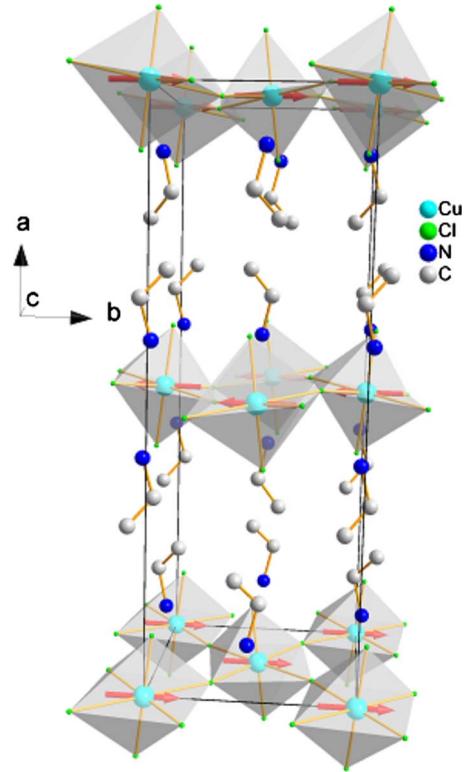

FIG. 1. (Color online) Schematic view of the orthorhombic crystal structure for $EA_2CuCl_4$ (300 K), the Jahn-Teller distorted copper chloride octahedra are highlighted. The ammonium heads ($NH_3^+$) of the organic cations form strong N-H···Cl hydrogen bonds to any of the eight halides that results in reorientational disorder of the bonding scheme; for simplicity the hydrogens of the $NH_3^+$ groups are not shown. The arrows at the copper sites are a graphical representation of the spin configuration involving adjacent $CuCl_4^{2-}$ layers at lower temperatures (see Ref. 34).

The temperature dependence of the magnetization for our sample (Fig. 2) supports a picture where ferromagnetic-like interactions appear to play a significant role. The data for our $EA_2CuCl_4$ sample are in agreement with the pioneering work of de Jongh *et al.*[33,34] who find a magnetic phase transition at $T_c=10.2$ K. The magnetization loops taken at different temperatures (Fig. 2, inset) show a quick saturation [reaching 1 $\mu_B$ per $Cu^{2+}$ ($S=1/2$) at 5 K], therefore supporting the presence of significant interactions below the magnetic ordering transition. It was claimed before that in a wide temperature range, the system shows characteristics of a two-dimensional Heisenberg ferromagnet with a dominant intralayer exchange coupling, $J/k_B=18.6$ K (Ref. 34) for the spins at the nearest-neigbor Cu sites (Fig. 1). However, this behavior is modified as one approaches $T_c$ from above because the interlayer coupling $J'$ becomes significant. Despite the small magnitude of the latter ($J'/J \approx -8 \times 10^{-4}$),[35] the susceptibility then resembles that of a system crossing over to a three-dimensional ordered AFM array of FM layers (Fig. 1). This is a consequence of the superposition of intraplane FM interaction and the very weak AFM-hole-orbital ordering (…$d_{x^2-z^2}$-$d_{y^2-z^2}$-$d_{x^2-z^2}$-$d_{y^2-z^2}$…) at the $Cu^{2+}$ ions that occupy antiferrodistortively arranged J-T elongated octahedra (within the $bc$ plane). This small antiferromagnetic coupling may be highly dependent on the interlayer distances and local lattice stresses that may additionally depend on materials' preparation procedure. Indeed the magnetic properties of this compound ($n=2$) were suggested to be strongly pressure dependent,[36] similarly to the pressure effects found in the case of the $n=1$ member of the family, i.e., $(CH_3NH_3)_2CuCl_4$.[37] Optical properties of $EA_2CuCl_4$ are also highly pressure dependent revealing a piezochromic effect.[38]

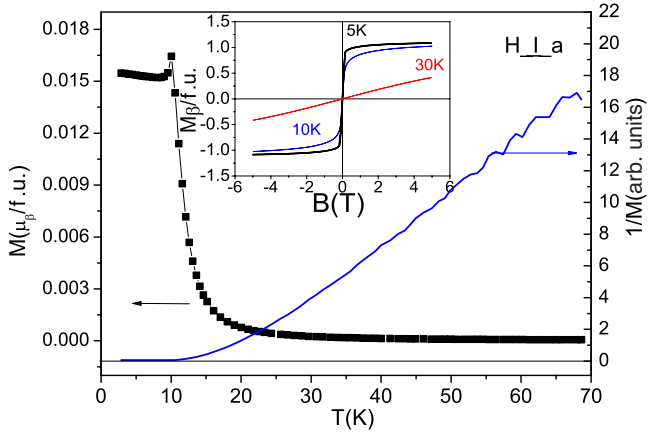

FIG. 2. (Color online) Magnetic field-cooled ($H=10$ Oe) magnetization (left scale) and its inverse plot (left scale). Inset: magnetization as a function of magnetic field at different temperatures.

Although $EA_2CuCl_4$ has been known for a long time, the ferroelectric and dielectric properties have never been reported in the low-temperature region; likewise, an electric polarization has not been studied. It is however documented that $EA_2CuCl_4$ undergoes a complicated series of phase transitions, mostly connected with changes in the arrangement of the alkylammonium chains.[39,40] Bearing in mind the low-temperature structural modifications here we report that as-grown alkylammonium copper chloride crystals develop a large electric polarization when the material is cooled below about 247 K. The temperature dependence of the real part of dielectric permittivity $\varepsilon'$ obtained on cooling shows an anomaly in the region of 200–260 K, with a peak around 247 K confirming a phase transition [Fig. 3(a)]. It is worth noting that both the broad shape and the maximum value of $\varepsilon'$ (at 247 K) are characteristic of improper ferroelectric phase transitions.

To a first approach, one could suggest that anomalies in $\varepsilon'_a$ on cooling are related to the same transition and the low-temperature wing appearing below $T_5$ is related to the influence of domain-wall dynamics. However, in the corresponding data obtained upon warming, we can resolve two closely separated maxima at around $T_4 \approx 232$ K and $T_5 \approx 247$ K that are likely to arise from different types of structural ordering, characteristic for this compound.[39,40] Analysis of the inverse dielectric susceptibility $\chi^{-1}$ as a function of temperature [inset to Fig. 3(a)] reveals a behavior that follows the Curie-Weiss law above $T_5$. Moreover, the linear extrapolation of $\chi^{-1}$ in the vicinity of the phase transition (temperature range of 250–255 K) intersects the $x$ axis at the same temperature, i.e., at 247 K both upon warming and cooling, thus pointing to a second-order type of phase transition at $T_5$. Under such circumstances the value $T_5=247$ K should be considered as an exact phase-transition temperature leading to polarization, while $T=232$ K represents an additional transition temperature in the sample. Therefore, the obvious deviation from Curie-Weiss behavior (between $T_5$ and $T_4$, as well as below $T_4$) and the fact that maximum in $\varepsilon'_a$ occurs at higher temperature on cooling (247 K) than on warming (232 K) are not features of the same transition (i.e., not a thermal hysteresis effect) but consequences of two closely situated phase transformations in the sample in the studied temperature range. It is necessary to note that the temperature dependences of the dielectric permittivity at cooling and warming show different slopes in the range of $T_4$ and 300 K. The main reason of such a behavior can be due to ferroelastic domain structure.[41] The effect of hysteresis arises from the fact that we crossover the $T_5$-$T_4$ temperature range and come back to room temperature so some ferroelastic domains (see Ref. 27) have not completely recovered to their paraelectric state.

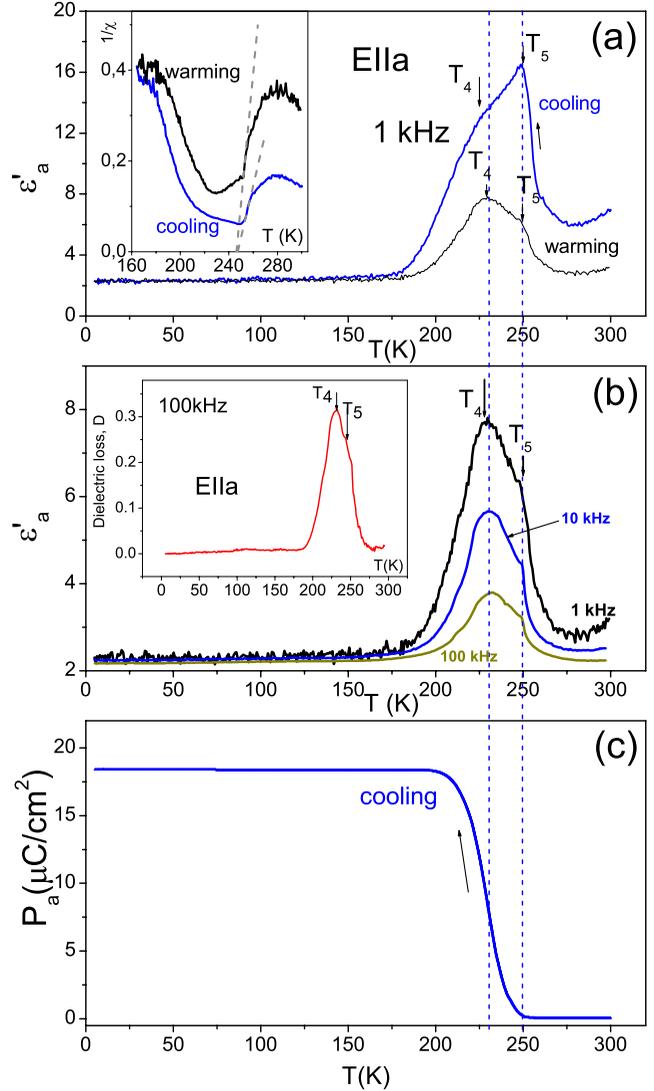

FIG. 3. (Color online) The temperature evolution of the dielectric permittivity, $\varepsilon'_a$, for a single-crystal $(C_2H_5NH_3)_2CuCl_4$ sample, (a) upon cooling and warming at $f=1$ kHz; inset: the inverse dielectric susceptibility in the vicinity of the phase transition and (b) the corresponding dielectric dispersion at various frequencies (on warming); inset: the dielectric loss, $D$, upon warming at $f=100$ kHz. (c) The temperature dependence of the spontaneous electric polarization, $P_a$, for $EA_2CuCl_4$ sample measured upon cooling.

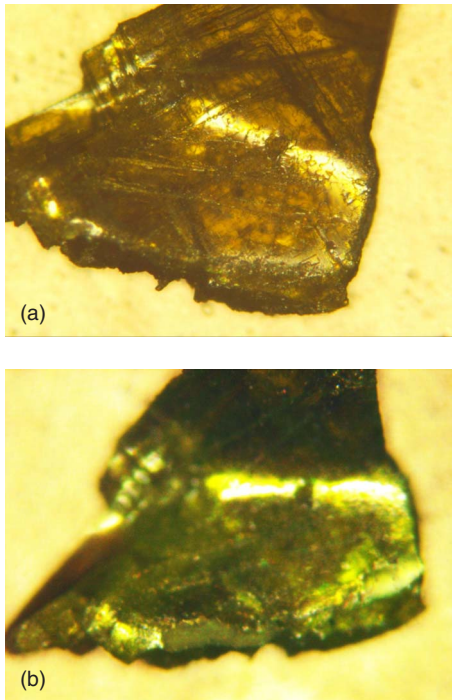

FIG. 4. (Color online) The color change in the sample (*bc* plane cut out) above [(a) yellow color] and below [(b) green color] the Curie point.

The temperatures at which the anomalies of the dielectric permittivity are observed ($T_4$ and $T_5$) are independent of the electric field frequency, $f$, confirming their relation to the phase transitions in EA$_2$CuCl$_4$ [Fig. 3(b)]. It is inferred that the anomalies in $\varepsilon'_a$ should be the result of order-disorder transitions connected with the reorientation of the alkylammonium groups as a whole. Organic chains flip among four equivalent orientations of the NH$_3$ group inside the cavity of the CuCl$_6$ octahdera which is a common feature in $(C_nH_{2n+1}NH_3)_2MCl_4$, pervoskites.[24,23,42] The anomalies corresponding to the above-mentioned phase transitions were also seen in the temperature dependence of dielectric loss, $D$ [inset Fig. 3(b)]. The assumption concerning the occurrence of the intermediate phase between $T_4$ and $T_5$ correlates fairly well with investigations involving ferroelastic domains in this system.[41] According to the latter report there are two types of domains above $T_5$, one of which vanishes below this temperature (247 K), whereas below $T_4$ all domains disappear. Under such circumstances the intermediate phase would be considered as ferroelastic-improper ferroelectric one. In any case this suggestion needs further confirmation by an independent experimental method, such as x-ray diffraction or optical ferroelectric domains observation under polarized light.

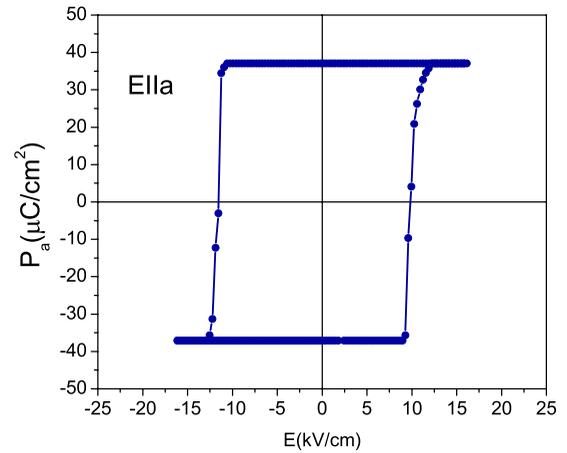

FIG. 5. (Color online) Compensated ferroelectric hysteresis loop for a $(C_2H_5NH_3)_2CuCl_4$ single crystal at 77 K.

In order to verify the polar response of EA$_2$CuCl$_4$ we studied the electric polarization by measuring the electric charge using automatic current integration with respect to time. Figure 3(c) shows the spontaneous electric polarization of the single-crystal sample measured upon cooling from 300 K to 5 K (3 K/min) without applied electric field. Indeed a large electric polarization, $P_a = 18 \ \mu C/cm^2$, is found to develop at $T < T_5$.

Besides the appearance of a spontaneous polarization the investigated crystals are characterized by a clear change in color in the ferroelectric phase: from yellow [Fig. 4(a)] at room temperature to light green at 77 K (not shown) and to dark green at approximately 100 K [Fig. 4(b)] (colors available online only). The effect of color change was already briefly mentioned in Ref. 19 and explained in terms structural distortion (near $T_4$), which modifies the optical energy gap.[38] Also in Ref. 28, changes in the absorption bands around the same temperature are reported.

Moreover, by demonstrating below that this compound is also ferroelectric, we open the rare class of magnetic multiferroics to the rich family of organic-inorganic perovskites where distinct model magnetic and electrically polar compounds can be discovered. The ferroelectric behavior is identified by the current integration with respect to time method, which after leakage compensation has provided a well-defined ferroelectric loop (Fig. 5) for our EA$_2$CuCl$_4$ single-crystal sample. Further efforts to measure ferroelectric properties at lower temperatures were unsuccessful most probably due to larger ferroelectric coercive force in the sample in this temperature region. It has to be noted that our experiments on magnetoelectric interactions in the sample (in a magnetic field $\pm 1$ T, at 9 K for $H_\parallel E$) did not show any noticeable coupling between magnetic and electric orders, in agreement with the different ordering temperatures of magnetic and electric subsystems. However, other configurations with respect to crystalline axis and electric field effect on magnetization remain to be tested. We have mentioned pre-viously that the temperature-dependent structural phase transitions in the EA$_2$CuCl$_4$ organic-inorganic perovskite are the result of order-disorder transitions connected with the alkylammonium group ordering,[23,24] however the final symmetry of the structure depends on other factors too, such as the J-T effect. Structural studies on these systems have shown before that at relatively high temperature the organic chains flip among four equivalent orientations of the NH$_3$ group inside the cavity of the $M$Cl$_6$ octahedra (Fig. 1). When the crystal is cooled down various orientational configurations are selected as the C$_n$H$_{2n+1}$NH$_3$ motion "freezes" gradually. This process is further refined due to the N-H⋯Cl hydrogen bonding, involving two different patterns, herein named as "aab" and "abb."[25] This terminology defines the bonding pattern, namely, with which type of chlorine the hydrogen bond of NH$_3$ group is formed—with terminal halogen (a) or with bridging (b) ones. We are of the opinion that this exact spontaneous reorientation of the hydrogen-bonding patterns and gradual ordering of the organic cation heads upon cooling the EA$_2$CuCl$_4$ to 247 K is the reason behind the observation of the anomaly in the dielectric permittivity and the measurement of a significant electric polarization. In addition to the exciting opportunities for multiferroic behavior in magnetic MOFs,[17] the present results demonstrate a proof of principle for the hydrogen bonding to help isolate magnetoelectrics from the diverse family of hybrid organic-inorganic perovskites.

We have provided compelling evidence that the $n=2$ member of the (C$_n$H$_{2n+1}$NH$_3$)$_2$CuCl$_4$ family is a magnetic multiferroic compound in which both electric polarization and dominant ferromagnetic interactions coexist. The (C$_2$H$_5$NH$_3$)$_2$CuCl$_4$ is a magnetic and ferroelectric compound that belongs to an emerging class of magnetic metal-organic multiferroics in which electric polarization is not induced by magnetic spin reorientation but ferroelectric and magnetic orders form independently. Considering the abundance of compounds with general formula (C$_n$H$_{2n+1}$NH$_3$)$_2$$MX_4$, where $M$ is a divalent metal, e.g., Mn$^{2+}$, Cd$^{2+}$, Fe$^{2+}$, and Cu$^{2+}$ and $X$ is a halogen, the present results warrants screening other similar magnetic materials where two functional subsystems are in action. The highly tunable organic building blocks in conjunction with 3$d$ metal layers in hybrid perovskites offer an approach where electric order induced by hydrogen bonding can coexist with (ferro)magnetic phases and help to enrich the rare family of room-temperature multiferroics.

B.K. acknowledges discussions with R. V. Lutciv and support from projects MELOIC (ANR-08-P196-36) of the "Agence Nationale de la Recherche" and BALISPIN (FF2008) from the "CNano Ile de France." The work was partially supported by the Ministry of Science and Education of Ukraine. A.L. acknowledges partial financial support to this project through the European Commissions' Marie Curie Transfer of Knowledge program NANOTAIL (Grant No. MTKD-CT-2006-042459).

---